\documentclass[%
reprint,
superscriptaddress,
bibnotes,twocolumn,
amsmath,amssymb,
aps,prl,
]{revtex4-2}

\usepackage{graphicx}
\usepackage{dcolumn}
\usepackage{bm}
\usepackage{hyperref}

\begin{document}
	
	
	\title{Thermal transport signatures of the excitonic transition and associated phonon softening in the layered chalcogenide Ta$_2$NiSe$_5$}

	\author{Yuan-Shan Zhang}
	\affiliation{Max Planck Institute for Solid State Research, 70569 Stuttgart, Germany}
	
	\author{Jan A. N. Bruin}%
	\email{j.bruin@fkf.mpg.de}
	\affiliation{Max Planck Institute for Solid State Research, 70569 Stuttgart, Germany}%
	
	\author{Yosuke Matsumoto}
	\affiliation{Max Planck Institute for Solid State Research, 70569 Stuttgart, Germany}%
	
	\author{Masahiko Isobe}
	\affiliation{Max Planck Institute for Solid State Research, 70569 Stuttgart, Germany}%
	
	\author{Hidenori Takagi}
	\email{h.takagi@fkf.mpg.de}
	\affiliation{Max Planck Institute for Solid State Research, 70569 Stuttgart, Germany}
	\affiliation{Institute for Functional Matter and Quantum Technologies, University of Stuttgart, 70569 Stuttgart, Germany}
	\affiliation{Department of Physics, University of Tokyo, 113-0033 Tokyo, Japan}
	

	\date{\today}
	
	\begin{abstract}
		The layered compound $\mathrm{Ta_2NiSe_5}$ is a quasi-one-dimensional and narrow-gap semiconductor, which is proposed to undergo a transition to an excitonic insulator at $T_\mathrm{c}=326$ K. We found a clear anomaly at $T_\mathrm{c}$ in the in-plane thermal conductivities both parallel ($\parallel$ $a$) and perpendicular ($\parallel$ $c$) to the one-dimensional chains, $\kappa_\mathrm{a}$ and $\kappa_\mathrm{c}$. While $\kappa_\mathrm{a}$ shows a rapid decrease below $T_\mathrm{c}$, $\kappa_\mathrm{c}$ shows a pronounced V-shaped suppression centered at $T_\mathrm{c}$. We argue that the decrease of $\kappa_\mathrm{a}$ represents the suppression of the quasiparticle contribution below $T_\mathrm{c}$ due to excitonic condensation. On the other hand, the V-shaped suppression of $\kappa_\mathrm{c}$ comes from the enhanced phonon scattering by soft phonons associated with the monoclinic transition with momentum $\mathbf{q}\parallel c$. The continued suppression of $\kappa_\mathrm{c}$ up to an extremely high temperature above $T_\mathrm{c}$ suggests the persistence of phonon softening likely coupled to electronic, presumably excitonic, fluctuations.
	\end{abstract}
	
	\maketitle

	The excitonic insulator (EI), in which bound electron-hole pairs form a condensate below a transition temperature $T_\mathrm{c}$, has been proposed theoretically more than 50 years ago \cite{jerome1967PRexcitonic}, but its realization in existing materials remains an active subject of investigation. The excitonic transition is anticipated in a nearly-zero-gap semiconductor when the exciton binding energy $E_\mathrm{b}$ exceeds the one-electron gap $E_\mathrm{g}$, or in a semimetal when the attractive interaction between electrons and holes is not screened out well. The realization in bulk materials has been discussed, for example, in $\mathrm{TmSe_{0.45}Te_{0.55}}$ \cite{wachter2004PRBpossibility} and $1T$-$\mathrm{TiSe_2}$ \cite{kogar2017Sciencesignatures}, but the evidence for excitonic condensation in these compounds has been far from convincing. The former is a $4f^{13}$ Mott-like system, not a band semiconductor/semimetal \cite{wachter2018AMPCexciton}, and the latter remains metallic and may suffer from strong screening of electron-hole interactions even below the proposed excitonic transition temperature \cite{di1976PRBelectronic}.
	
\begin{figure}
		\includegraphics[scale=1,keepaspectratio]{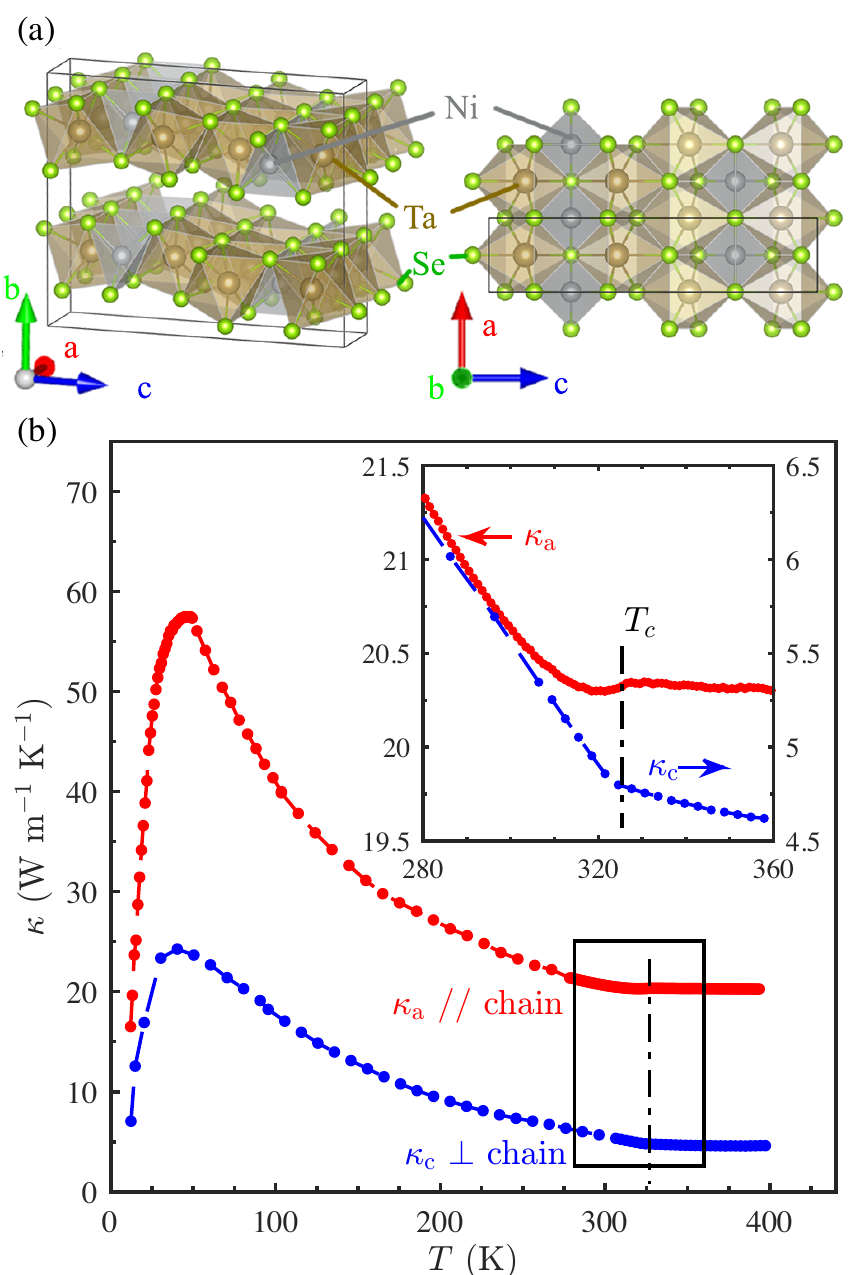}
		\caption{(a) Crystal structure of $\mathrm{Ta_2NiSe_5}$ showing the stack of layers along the $b$ axis (left) and a single layer in the $a$-$c$ plane consisting of an array of two chains of Ta octahedrally coordinated with Se and one chain of Ni in between, tetrahedrally coordinated with Se (right) \cite{momma2011JACvesta}. (b) Thermal conductivities $\kappa_{\mathrm{a}}$ with heat current parallel to the chains ($\parallel$ $a$; red) and $\kappa_{\mathrm{c}}$ with heat current perpendicular to the chains ($\parallel$ $c$; blue) as a function of temperature. The dashed black line indicates $T_{\mathrm{c}}=326$ K. Inset: enlarged view of $\kappa_{\mathrm{a}}$ and $\kappa_{\mathrm{c}}$ near $T_{\mathrm{c}}$.}
		\label{fig1}
	\end{figure}

	Very recently, $\mathrm{Ta_2NiSe_5}$ has emerged as a leading candidate for an excitonic insulator \cite{wakisaka2009PRLexcitonic, Kim_ACSN_2016, lu2017zero, larkin2017PRBgiant, Mor_PRL_2017, werdehausen2018SAcoherent, Seo_SR_2018, Okazaki_NC_2018,fukutani2019PRLelectrical, chen2020PRBdoping, volkov2020critical, kim2020direct, he2020tunnelingtipinduced}. It is an almost-zero-energy- and direct-gap semiconductor with Ta $5d$ conduction bands and Ni $3d$ valence bands \cite{lu2017zero}. It shows a transition to an insulator at  $T_\mathrm{c}=326$ K, which was proposed to be an excitonic transition. The crystal structure of $\mathrm{Ta_2NiSe_5}$ is layered, and as shown in Fig. \ref{fig1}(a), each layer consists of an array of Ta and Ni chains running along the $a$ axis, which renders the system electronically quasi-one-dimensional and gives rise to spatially isolated electron and hole chains. The one dimensionality is known to enhance the exciton binding energy, which makes $\mathrm{Ta_2NiSe_5}$ an ideal platform for the search for an excitonic insulator. Well below $T_\mathrm{c}$, an optical gap of 0.16 eV develops, which is close to the exciton binding energy of 0.25 eV experimentally observed in the optical conductivity spectrum of the sister compound $\mathrm{Ta_2NiS_5}$. $\mathrm{Ta_2NiS_5}$ has a much larger gap of 0.6 eV and exhibits no signatures of a phase transition \cite{lu2017zero, larkin2017PRBgiant}. This strongly suggests the excitonic origin of the insulating gap in $\mathrm{Ta_2NiSe_5}$ below $T_\mathrm{c}$. As a function of the one-electron gap $E_\mathrm{g}$ controlled by pressure and chemical substitution, the transition temperature $T_\mathrm{c}$ shows a dome-shaped behavior centered at $E_\mathrm{g}\approx0$, corresponding to pure $\mathrm{Ta_2NiSe_5}$ at ambient pressure \cite{lu2017zero}. This is fully consistent with what is expected for the canonical excitonic insulator \cite{jerome1967PRexcitonic}.
	
	The aforementioned experimental results clearly point to the excitonic character of the transition. Nevertheless, it has become increasingly clear that the transition cannot be purely excitonic and that the actual situation is more complicated \cite{watson2020PRRband,subedi2020PRMorthorhombic,baldini2020arxiv_spontaneous}, which makes the physics of the putative excitonic transition even more challenging and attractive. The putative excitonic transition at $T_\mathrm{c}=326$ K is accompanied by a structural phase transition from an orthorhombic to a monoclinic structure \cite{di1986JLCMphysical, nakano2018IUXRJpressure}. In the monoclinic phase, the $\mathrm{TaSe_2}$ chain-ladder unit distorts, such that the Ta chain and the surrounding Se ladders move along the $a$ axis, but in opposite directions \cite{nakano2018PRBantiferroelectric}. The movement of two neighboring $\mathrm{TaSe_2}$ chain-ladder units surrounding the Ni chain is asymmetric with respect to the Ni chain. This shear-type distortion with antiferroelectric-type displacement of the Ta and the Se chains does not bring about any superstructure (i.e., $\mathbf{q}=0$) but allows for the hybridization of Ta conduction bands and Ni valence bands around the $\Gamma$ point. The resultant hybridization gap could account for at least a part of the insulating gap and should be one of the important ingredients of the transition at 326 K. A symmetry analysis demonstrated that the excitonic order parameter must be accompanied by the monoclinic distortion and hence the admixture of the excitonic gap and the hybridization gap is unavoidable \cite{mazza2020PRLnature}.
	
	In this study, we measure the in-plane anisotropic thermal conductivities $\kappa_\mathrm{a}(T)$ along the chains ($\parallel$ $a$) and $\kappa_{\mathrm{c}}(T)$ perpendicular to the chains ($\parallel$ $c$) of $\mathrm{Ta_2NiSe_5}$ single crystals, in order to unveil the quasiparticle thermal conductivity of the putative excitonic insulator and the phonon transport around the monoclinic transition. In excitonic insulators, quasiparticle transport analogous to Bardeen-Cooper-Schrieffer (BCS) superconductors may be expected \cite{jerome1967PRexcitonic,tinkham2004CCintroduction}. Thermal conductivity is an especially useful probe of the quasiparticle transport in an excitonic insulator as the cancelation between the electron and hole quasiparticles does not take place. In addition, thermal conductivity has a large phonon contribution and is sensitive to soft phonons associated with the monoclinic transition, which may give us hints to decipher the role of the lattice. We show that $\kappa_{\mathrm{a}}(T)$ along the chains is rapidly suppressed below $T_\mathrm{c}$ and that it comes from the decrease in number of quasiparticles (electrons and holes) below $T_\mathrm{c}$, which is expected for a canonical excitonic insulator \cite{zittartz1968PRtransport}. In contrast, $\kappa_{\mathrm{c}}$ perpendicular to the chains shows a V-shaped suppression due to enhanced scattering by soft phonons associated with the orthorhombic-to-monoclinic transition with momentum $\mathbf{q}$ perpendicular to the chains ($\parallel$ $c$). The strong suppression of $\kappa_{\mathrm{c}}$ persists up to more than 100 K above $T_\mathrm{c}$, which suggests a robust phonon softening likely coupled to excitonic (and hybridization-gap) fluctuations.
	
	Needle-shaped single crystals of $\mathrm{Ta_2NiSe_5}$ with typical size of $\mathrm{4~mm\times1~mm\times0.1~mm}$ were grown by chemical vapor transport \cite{lu2017zero,yuanshan2020signatureSI}, with their longest dimension along the Ta/Ni chains ($a$ axis) and their shortest dimension along the interlayer $b$ axis. Rectangular-shaped samples with their long axis along the $a$ and $c$ axes were cut out from the single crystals. The electrical resistivity of the samples was measured by a conventional four-probe technique, while thermal conductivity measurements along the same axes were performed via a ``heat-pipe'' method \cite{allen1994PRBthermal,wakeham2011NCgross}. The temperature gradients were detected with a differential type-E thermocouple. A transition was observed at $T_\mathrm{c}=326$ K as a clear kink in the electrical resistivity along the chain direction ($\rho_{\mathrm{a}}$), as reported previously \cite{lu2017zero,yuanshan2020signatureSI}.
	
	As shown in Fig. \ref{fig1}(b), $\kappa_{\mathrm{a}}(T)$ along the chains ($\parallel$ $a$) is a factor of 3-4 larger than $\kappa_{\mathrm{c}}$ perpendicular to the chains ($\parallel$ $c$). With increasing temperature from 10 K, both $\kappa_{\mathrm{a}}$ and $\kappa_{\mathrm{c}}$ increase and show a peak around 30-40 K, followed by a gradual decrease that continues above room temperature. This behavior is common for a crystalline solid with phonon-dominated thermal conductivity \cite{berman1976CPthermal,tritt2005SSBMthermal}.
	
	At the transition temperature $T_\mathrm{c}=326$ K, we observe a clear anomaly both in $\kappa_{\mathrm{a}}(T)$ and $\kappa_{\mathrm{c}}(T)$. The anomaly at $T_\mathrm{c}$, however, shows up in a contrasted manner between the two directions $a$ and $c$, which can be seen more clearly in the enlarged plot around $T_\mathrm{c}$ in the inset of Fig. \ref{fig1}(b). $\kappa_{\mathrm{a}}(T)$ parallel to the chains shows a shoulder structure with a well-defined kink at $T_\mathrm{c}$, followed by an abrupt drop below $T_\mathrm{c}$. Upon further lowering of the temperature, $\kappa_{\mathrm{a}}(T)$ recovers and begins increasing below $\sim$315 K. In contrast, $\kappa_{\mathrm{c}}(T)$ shows a discontinuous change of slope at $T_\mathrm{c}$ from a weak increase to a very rapid increase with decreasing temperature, which can be viewed as a V-shaped suppression from the $1/T$-like increase expected for phonon-dominant, high-temperature thermal transport \cite{berman1976CPthermal}. The contrast is even clearer in the temperature derivatives, $\partial\kappa_{\mathrm{a}}/\partial T$ and $\partial\kappa_{\mathrm{c}}/\partial T$ \cite{yuanshan2020signatureSI}. We argue that the contrasted behaviors of the anomaly at $T_\mathrm{c}$ between $\kappa_{\mathrm{a}}(T)$ and $\kappa_{\mathrm{c}}(T)$ come from the predominant electronic and phonon contributions in the temperature dependence of $\kappa_{\mathrm{a}}(T)$ and $\kappa_{\mathrm{c}}(T)$, respectively, near $T_\mathrm{c}$, which allows us to discuss the critical behaviors of quasiparticles (electrons and holes) and phonons around $T_\mathrm{c}$ separately.
	
\begin{figure}
		
		\includegraphics[scale=1]{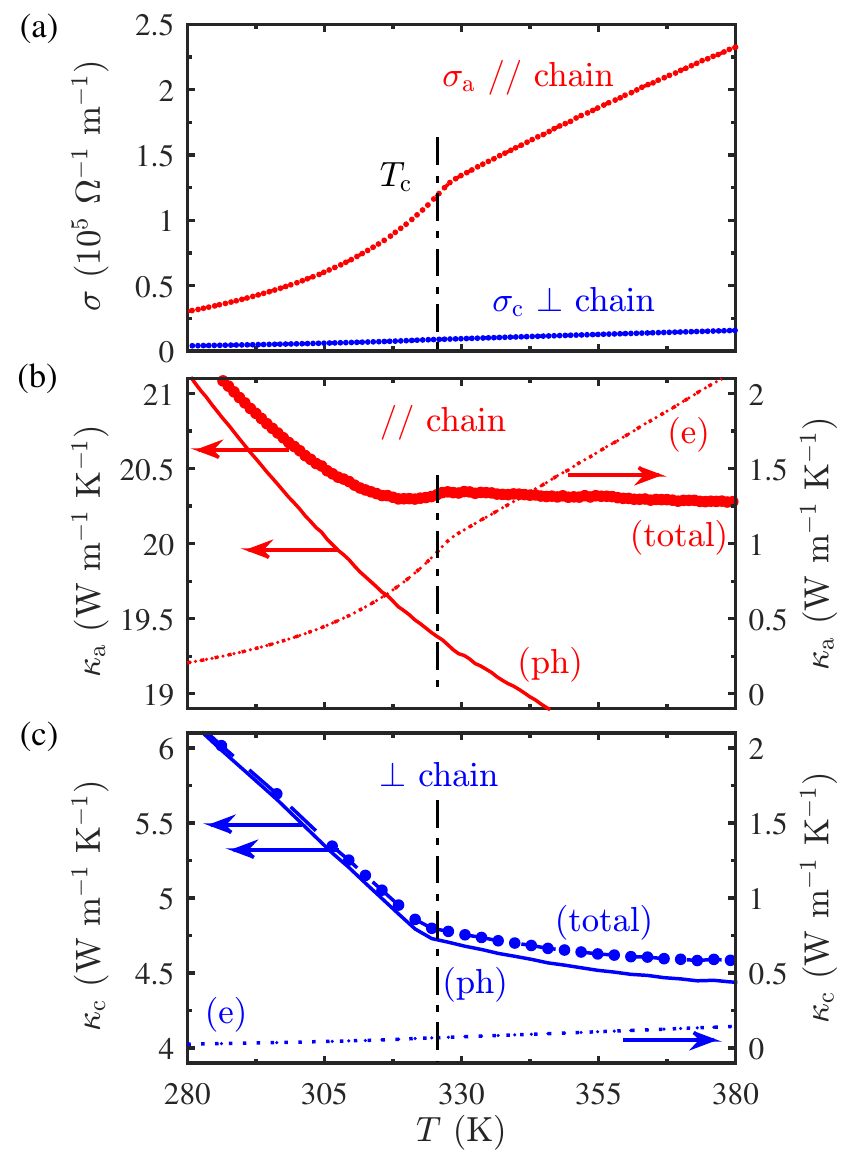}
		\caption{Separation of electron and phonon contributions to the thermal conductivities $\kappa_{\mathrm{a}}$ with heat current parallel to the chains ($\parallel$ $a$; red) and $\kappa_{\mathrm{c}}$ with heat current perpendicular to the chains ($\parallel$ $c$; blue) in Fig. \ref{fig1}, using the Wiedemann-Franz law. (a) The temperature-dependent electrical conductivities, $\sigma_{\mathrm{a}}(T)$ (red) parallel to the chains and $\sigma_{\mathrm{c}}(T)$ (blue) perpendicular to the chains. (b) The total, electron, and phonon thermal conductivities along the $a$ axis, parallel to the chains, denoted by $\kappa_{\mathrm{a}}$ (circles), $\kappa_{\mathrm{a}}^{(\mathrm{e})}$ (broken), and $\kappa_{\mathrm{a}}^{(\mathrm{ph})}$ (solid), respectively. (c) The total, electron, and phonon thermal conductivities along the $c$ axis, perpendicular to the chains, denoted $\kappa_{\mathrm{c}}$ (circles), $\kappa_{\mathrm{c}}^{(\mathrm{e})}$ (broken), and $\kappa_{\mathrm{c}}^{(\mathrm{ph})}$ (solid), respectively.  }
		\label{fig2}
	\end{figure}

	The measured thermal conductivity $\kappa$ comprises of contributions from electrons, $\kappa^{(\mathrm{e})}$, and from phonons, $\kappa^{(\mathrm{ph})}$. Around $T_\mathrm{c}$, $\kappa^{(\mathrm{e})}$ can be estimated from the electrical conductivity $\sigma$ using the Wiedemann-Franz (W-F) law. Strictly speaking, the W-F law is valid only at $T=0$ K. Nevertheless, it is also a realistic approximation at finite temperatures, where large-angle scattering processes dominate \cite{berman1976CPthermal,tritt2005SSBMthermal}. $\kappa^{(\mathrm{ph})}$ can be estimated as the remnant $\kappa-\kappa^{(\mathrm{e})}$. As shown in Fig. \ref{fig2}(a), the electrical conductivity along the chains, $\sigma_{\mathrm{a}}$, is an order of magnitude larger than that perpendicular to the chains, $\sigma_{\mathrm{c}}$, reflecting the quasi-one-dimensional band structure of $\mathrm{Ta_2NiSe_5}$. In addition, the anomaly at $T_\mathrm{c}$ is much more pronounced in $\sigma_{\mathrm{a}}(T)$ as compared with $\sigma_{\mathrm{c}}(T)$. A much larger electronic contribution to the temperature dependence of $\kappa_{\mathrm{a}}(T)$ than $\kappa_{\mathrm{c}}(T)$ is thus expected.
	
	As shown in Fig. \ref{fig2}(b), the electronic thermal conductivity along the chains $\kappa_{\mathrm{a}}^{(\mathrm{e})}(T)$, estimated from $\sigma_{\mathrm{a}}$, is only $\sim$5$\%$ of the total $\kappa_{\mathrm{a}}(T)$, but fully accounts for the shoulder structure in $\kappa_{\mathrm{a}}(T)$ with the rapid decrease below $T_\mathrm{c}$. On the other hand, the phonon thermal conductivity $\kappa_{\mathrm{a}}^{(\mathrm{ph})}(T)$, estimated by subtracting $\kappa_{\mathrm{a}}^{(\mathrm{e})}(T)$ from $\kappa_{\mathrm{a}}(T)$, is surprisingly featureless and shows a monotonic increase through $T_\mathrm{c}$, indicating that the anomaly at $T_\mathrm{c}$ in $\kappa_{\mathrm{a}}(T)$ is purely electronic in origin. The rapid decrease below $T_\mathrm{c}$ therefore reflects the decrease in number of quasiparticles, electrons and holes, in analogy with the prototypical behavior of thermal conductivity in a superconductor \cite{satterthwaite1962PRthermal,ambegaokar1965PRcalculation}. We therefore conclude that the quasiparticle transport of a putative excitonic insulator is captured by thermal transport. 
	
	In superconductors, the thermal conductivity often does not show a rapid decrease right below $T_\mathrm{c}$, the prototypical behavior, but instead shows an increase. This can be ascribed to the suppression of scattering of electrons and phonons by electrons due to the opening of an electronic gap below $T_{\mathrm{c}}$. In certain strongly correlated superconductors with anisotropic pairing, such as high-$T_{\mathrm{c}}$ cuprates \cite{yu1992PRLthermal}, the enhancement of quasiparticle lifetime due to the suppression of strong electron-electron scattering overcomes the decrease of quasiparticle density and thereby results in an increase of $\kappa$ right below $T_{\mathrm{c}}$. In certain conventional electron-phonon superconductors, the increase of phonon mean free path due to the suppression of strong scattering of phonons by electrons may also result in the enhancement of phonon thermal conductivity below $T_{\mathrm{c}}$, which brings about the increase of $\kappa$ right below $T_{\mathrm{c}}$ \cite{carlson1970PRLanomalous}. The observation of a rapid decrease in $\kappa_{\mathrm{a}}(T)$ below $T_{\mathrm{c}}$ in $\mathrm{Ta_2NiSe_5}$ thus indicates that the scattering of phonons and electrons with momentum parallel to the chains by electrons should be reasonably weak as compared with other dominant scattering mechanisms, such as phonon-phonon scattering for phonon transport.
	
\begin{figure}	
		\includegraphics[scale=1]{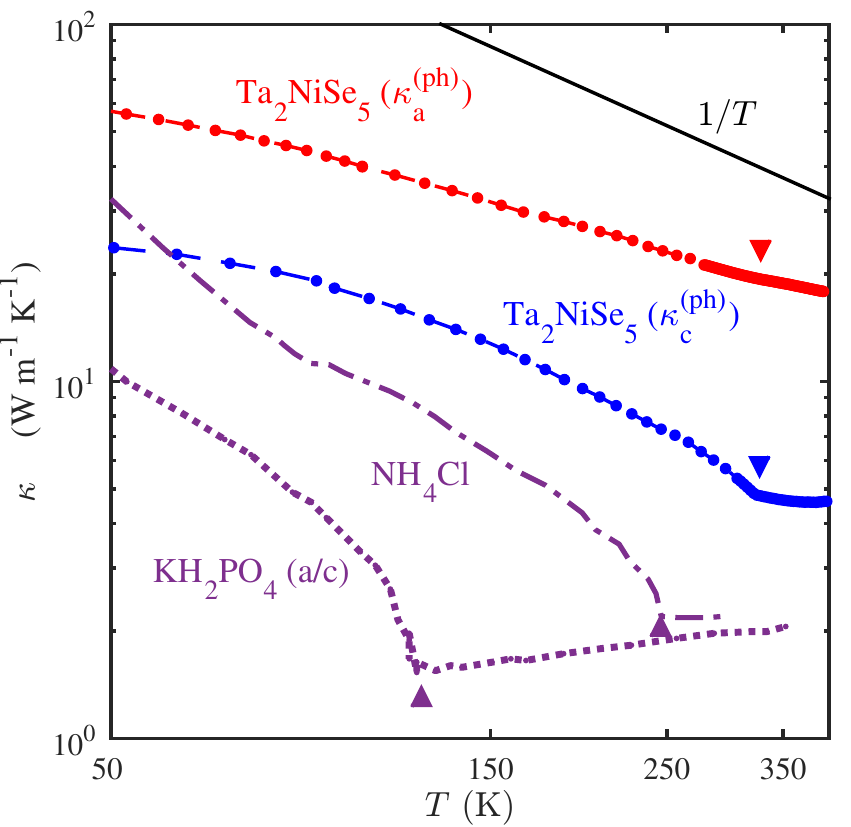}
		\caption{Comparison of temperature-dependent thermal conductivities $\kappa(T)$ for $\mathrm{Ta_2NiSe_5}$ ($\parallel$ $a$ red; $\parallel$ $c$ blue) with typical order-disorder-type ferroelectrics, $\mathrm{NH_4Cl}$ (reproduced from \cite{bausch1969PLAthermal}) and $\mathrm{KH_2PO_4}$ (reproduced from \cite{suemune1967JPSJthermal}). The solid black line is a guide line for the $1/T$ power law.}
		\label{fig3}
	\end{figure}

	As $\sigma_{\mathrm{c}}$ is one order of magnitude smaller and exhibits much weaker structure at $T_{\mathrm{c}}$ than $\sigma_{\mathrm{a}}$, the electronic thermal conductivity $\kappa_{\mathrm{c}}^{\mathrm{(e)}}(T)$ perpendicular to the chain direction is as small as $\sim$1$\%$ of the total $\kappa_{\mathrm{c}}(T)$ in magnitude, and, more importantly, does not contribute appreciably to the kink-like anomaly at $T_{\mathrm{c}}$ in the temperature dependence of $\kappa_{\mathrm{c}}(T)$, as seen in Fig. \ref{fig2}(c). In contrast to $\kappa_{\mathrm{a}}(T)$ along the chains, $\kappa_{\mathrm{c}}(T)$ around $T_{\mathrm{c}}$ is dominated by $\kappa_{\mathrm{c}}^{(\mathrm{ph})}(T)$ and almost purely phononic in origin. The comparison of $\kappa_{\mathrm{c}}^{(\mathrm{ph})}(T)$ and  $\kappa_{\mathrm{a}}^{(\mathrm{ph})}(T)$ reveals a substantial anisotropy not only in electron transport but also in phonon transport between the directions parallel and perpendicular to the chains.  As clearly seen in Fig. \ref{fig3}, while a $1/T$-like monotonous decrease with featureless behavior around $T_{\mathrm{c}}$ is observed in $\kappa_{\mathrm{a}}^{(\mathrm{ph})}(T)$, a clear V-shaped suppression from the $1/T$-like decrease centered at $T_{\mathrm{c}}$ is seen in $\kappa_{\mathrm{c}}^{(\mathrm{ph})}(T)$. The suppression is broad in temperature range and appears to be asymmetric above and below $T_{\mathrm{c}}$. While  $\kappa_{\mathrm{c}}^{(\mathrm{ph})}(T)$ shows a rapid increase and crossover to $1/T$-like behavior with decreasing temperatures right below $T_{\mathrm{c}}$, it is almost flat up to at least 400 K with increasing temperature from $T_{\mathrm{c}}$, implying the persistence of strong suppression over a wide temperature range above $T_{\mathrm{c}}$.
	
	Phonon thermal conductivity is generally expressed as $\kappa^{\mathrm{(ph)}}=\frac{1}{3}C^{\mathrm{(ph)}}v^{\mathrm{(ph)}}l^{(\mathrm{ph})}$, where $C^{\mathrm{(ph)}}$, $v^{\mathrm{(ph)}}$, and $l^{(\mathrm{ph})}$ are the specific heat, the velocity, and the mean free path of phonons \cite{berman1976CPthermal}. The temperature dependence of the total heat capacity $C$ for $\mathrm{Ta_2NiSe_5}$ around $T_{\mathrm{c}}=326$ K is rather small and the small anomaly at $T_{\mathrm{c}}$ is highly likely to be purely electronic in origin \cite{lu2017zero}. The temperature dependence of $C^{\mathrm{(ph)}}$ can thus be ignored in discussing the origin of the V-shaped suppression of $\kappa_{\mathrm{c}}^{\mathrm{(ph)}}(T)$. The suppression of $\kappa_{\mathrm{c}}^{\mathrm{(ph)}}(T)$ should originate instead from the reduction of $v^{\mathrm{(ph)}}$ and/or $l^{(\mathrm{ph})}$ around $T_{\mathrm{c}}$. As it occurs only for phonons with momentum $\mathbf{q}\parallel c$, it is natural to ascribe its origin to the reported softening of the acoustic shear-mode phonon with $\mathbf{q}\parallel c$, which corresponds to the monoclinic distortion of lattice frozen at the excitonic transition $T_{\mathrm{c}}=326$ K \cite{nakano2018PRBantiferroelectric}. As pronounced softening occurs only for one specific acoustic phonon mode, the reduction of $v^{\mathrm{(ph)}}$ is unlikely to account for the large suppression of $\kappa_{\mathrm{c}}^{\mathrm{(ph)}}(T)$ at $T_{\mathrm{c}}$ [Fig. \ref{fig2} (c)]. The primary origin of the suppression of $\kappa_{\mathrm{c}}^{\mathrm{(ph)}}(T)$ is highly likely the increased scattering of phonons, namely the reduction of $l^{(\mathrm{ph})}$, by the soft acoustic shear-mode phonon. The strong suppression up to at least 400 K above $T_{\mathrm{c}}$ should then imply the robust softening of the shear mode acoustic phonon above $T_{\mathrm{c}}$, which is indeed quite consistent with the reported phonon dispersion relationship of $\mathrm{Ta_2NiSe_5}$ measured by inelastic X-ray scattering up to 400 K \cite{nakano2018PRBantiferroelectric}.
	
	We should point out that the asymmetric V-shaped suppression of $\kappa_{\mathrm{c}}^{\mathrm{(ph)}}(T)$, more robust above $T_{\mathrm{c}}$, is analogous to those observed around the order-disorder-type ferroelectric transition in $\mathrm{KH_2PO_4}$ and $\mathrm{NH_4Cl}$, as shown in Fig. \ref{fig3}. The order-disorder-type ferroelectrics have extended ferroelectric fluctuations above $T_{\mathrm{c}}$, where well-developed but fluctuating local dipoles give rise to soft local phonons and enhanced phonon scattering over a wide temperature range, and hence suppress $\kappa(T)$ strongly over a wide temperature range above $T_{\mathrm{c}}$ \cite{lines2001OXFprinciples,suemune1967JPSJthermal,bausch1969PLAthermal,strukov1994PTMJheat}. It is tempting to infer that, in analogy with order-disorder type ferroelectrics, large and local excitonic and/or hybridization gap ($\mathbf{q}=0$ charge density wave) fluctuations may be present above $T_\mathrm{c}$ in $\mathrm{Ta_2NiSe_5}$. The electronic fluctuations are inherently and strongly coupled to the shear mode phonon, which may give rise to the extended phonon softening above $T_c$ and may suppress $\kappa_{\mathrm{c}}^{\mathrm{(ph)}}(T)$ over a wide temperature range. Such giant excitonic and/or hybridization gap fluctuations may be inferred from the temperature dependence of the bulk magnetic susceptibility $\chi(T)$ of $\mathrm{Ta_2NiSe_5}$ \cite{di1986JLCMphysical}, where $\chi(T)$ at temperatures well above $T_{\mathrm{c}}$ shows a pronounced downturn upon lower temperatures towards $T_{\mathrm{c}}$, indicative of such fluctuations.

	In summary, we have discovered an extremely anisotropic behavior of the thermal conductivity around the transition from an almost-zero gap semiconductor to an insulator in a putative excitonic insulator, the layered chalcogenide $\mathrm{Ta_2NiSe_5}$. The distinct temperature-dependent in-plane $\kappa_{\mathrm{a}}(T)$ and $\kappa_{\mathrm{c}}(T)$, parallel and perpendicular to the Ta and Ni chains, respectively, allow us to unveil the behaviors of quasiparticles (electrons and holes) and phonons around the transition temperature $T_{\mathrm{c}}=326$ K. The rapid suppression in the number of quasiparticles below $T_{\mathrm{c}}$, analogous to superconductors, is captured in $\kappa_{\mathrm{a}}(T)$, due to the quasi-one-dimensional electronic structure and the $\mathbf{q} \parallel a$ phonons being insensitive to the transition. The soft shear-mode acoustic phonons with momentum $\mathbf{q} \parallel c$, which corresponds to the monoclinic structural distortion below $T_{\mathrm{c}}$, manifest themselves only in $\kappa_{\mathrm{c}}(T)$ perpendicular to the chains as a pronounced suppression of dominant phonon thermal transport. The suppression of $\kappa_{\mathrm{c}}$ extends up to high temperatures well above $T_{\mathrm{c}}$, indicating the persistence of phonon softening over a wide temperature range. This may suggest the presence of giant excitonic and/or hybridization gap fluctuations above $T_{\mathrm{c}}$ and the strong electron-lattice coupling for the monoclinic distortion in $\mathrm{Ta_2NiSe_5}$.

	\begin{acknowledgments}
		This work is supported in part by the Alexander von Humboldt foundation. We thank L. Dorner-Finkbeiner and C. Bush for experimental support in crystal growth and characterization, A. Bangura for suggesting the “heat-pipe” technique, and D. Huang for useful discussions.
	\end{acknowledgments}


	%

\end{document}